\documentclass[
reprint,
superscriptaddress,
amsmath,amssymb,
aps,
prx,
longbibliography,
]{revtex4-2}

\usepackage{graphicx}
\usepackage{dcolumn}
\usepackage{bm}

\usepackage{tcolorbox}
\usepackage{booktabs,tabularx,array}
\usepackage[colorlinks=true,linkcolor=blue,citecolor=blue,urlcolor=blue]{hyperref}

\newcolumntype{Y}{>{\centering\arraybackslash}X}

\begin{document}

\preprint{APS/123-QED}

\title{\textbf{Test Particle Study of EDI Driven Electron Transport in a Hall Thruster Using PIC Derived Electric Fields} 
}%

\author{Zhongping Zhao}%
\author{Kunpeng Zhong}
\author{Yinjian Zhao}
 \email{Contact author: zhaoyinjian@hit.edu.cn}
 
\affiliation{School of Energy Science and Engineering, Harbin Institute of Technology, Harbin 150001, People’s Republic of China}

\date{\today}

\begin{abstract}
Electron transport in a Hall thruster is investigated at the test particle level using prescribed electric fields derived from the electron drift instability (EDI) resolving three dimensional particle-in-cell (PIC) simulation. Four primary electric field configurations are considered: static averaged field, full PIC field, proper orthogonal decomposition (POD) reconstructed field, and a simplified analytical $E_y$ field.
Two additional control cases, the magnetic-field-only case and the denoised PIC field, are also included.
Transport statistics show that the averaged field produces only weak axial cross field transport. In contrast, the full PIC field produces clear fluctuation driven transport, characterized by pronounced negative axial displacement, enhanced channel entry, finite channel residence, particle energization, and appreciable anode directed loss. POD analysis shows that the transport relevant EDI electric field structures are distributed over multiple coupled modes, and that a moderate truncation order of approximately $20$ modes is required to recover the main transport signatures.
The analytical $E_y$ model separately examines the influence of azimuthal electric field fluctuations on axial electron transport and shows that fluctuation amplitude is a primary determinant.

\end{abstract}

\maketitle


\section{Introduction}

Hall thrusters (HTs) are mature plasma electric propulsion devices in which an axial electric field accelerates ions to generate thrust, while an externally applied, predominantly radial magnetic field magnetizes electrons and induces an azimuthal $\mathbf{E}\times\mathbf{B}$ drift \cite{goebel2023fundamentals,lafleur2017characteristics}. However, suppressing electron transport along the electric field direction by means of a perpendicular magnetic field is far from trivial; numerous studies have shown that the cross field electron mobility remains anomalously high and cannot be fully accounted for by classical collisional diffusion theories based on standard electron-neutral or electron-ion collisions \cite{boeuf2023physics,boeuf2017tutorial,boeuf2018b,kaganovich2020physics,lafleur2016theory1}. Accordingly, the electron drift instability (EDI), a kinetic instability often interpreted as arising from the coupling between electron Bernstein modes and ion acoustic waves, has been proposed as a key mechanism underlying anomalous electron transport \cite{PhysRevLett.130.115101,PhysRevE.108.065204}, whereby the azimuthal drift of magnetized electrons relative to essentially unmagnetized ions excites electrostatic fluctuations with characteristic millimeter-scale wavelengths and MHz-scale frequencies, thereby enhancing cross field electron mobility \cite{lafleur2016theory2,gary1970longitudinal1,gary1970longitudinal2,PST-2025-0248}.

Given the kinetic nature of the EDI, PIC simulations provide a natural framework for resolving the associated wave--particle dynamics and assessing their contribution to cross field electron transport. One dimensional azimuthal PIC simulations have isolated the basic EDI mechanism and attributed the enhanced transport to correlations between plasma density fluctuations and azimuthal electric field fluctuations \cite{lafleur2016theory1,smolyakov2020anomalous,asadi2019numerical}. Two dimensional axial--azimuthal PIC simulations have further mapped the axial evolution of the azimuthal $E_y$ spectrum. These studies revealed a transition of the dominant modes from short wavelength, high frequency structures upstream of the magnetic field maximum $(\lambda \approx 0.5\,\mathrm{mm},\, f \approx 5\,\mathrm{MHz})$ to longer wavelength, lower frequency structures downstream $(\lambda \approx 2\,\mathrm{mm},\, f \approx 3\,\mathrm{MHz})$, with the upstream spectrum exhibiting ion acoustic like dispersion \cite{boeuf2018b,charoy20192d,PST-2026-0036}. Despite these advances, reduced dimensional PIC simulations remain projections of the full three dimensional EDI dynamics. More recent 3D PIC simulations have shown that resolving the radial direction, approximately parallel to $\mathbf{B}$, reveals oblique radial--azimuthal $E_y$ structures, additional fluctuations possibly associated with the modified two-stream instability (MTSI) or longer wavelength modes, and reduced fluctuation driven transport in channel and near plume simulations compared with analogous 2D cases \cite{villafana20233d,taccogna2018three,chen2025influence}. Although 3D PIC simulations can resolve EDI dynamics with high fidelity while advancing all macroparticles self-consistently, their computational and storage costs remain substantial \cite{zhao2026review,ZHONG2026131809,
zhao2025effect,
chen2025coupling,chen2025influence}.

The test particle method provides a complementary diagnostic approach. In this method, selected particles are advanced in prescribed electric and magnetic fields, with collisional effects optionally included \cite{smith2010single,marini2017single}. The particles are treated as passive tracers: they respond to the imposed fields but do not deposit charge and therefore do not modify the field evolution. This passive formulation makes it possible to archive complete phase space histories for all test particles and to examine individual trajectories in detail. The method is therefore well suited for connecting EDI resolving electric field structures to microscopic electron dynamics and anomalous transport.

The test particle approach has previously been used to isolate specific mechanisms of electron transport in HTs. Perez-Luna \textit{et al.}~\cite{perez2008electron} showed that an imposed azimuthal potential perturbation can drive electron trajectories toward the anode and produce mean axial velocities of the same order as experimental estimates, although the perturbation was prescribed as a static sinusoidal mode. Coche and Garrigues \cite{coche2011study} later used a test particle Monte Carlo model driven by force fields from a 2D axial--azimuthal PIC simulation and showed that replacing the full fluctuating field with a single mode wave underestimates the electron mobility. Smith and Cappelli \cite{smith2007numerical,smith2010single,smith2006investigation} performed 3D single particle simulations of cathode emitted electrons in the near field of a HT using prescribed static electric and magnetic fields. Their calculations showed that nonuniform near field $\mathbf{E}$ and $\mathbf{B}$ structures, together with front face surface collisions, can yield a channel-to-plume current ratio of order $0.1$ and electron lifetimes of order $120\,\mathrm{ns}$ in a $0.3\,\mathrm{m}$ axial domain. Their work established single particle simulations as a useful microscopic diagnostic of cross field electron motion. However, their study focused on transport from the near field cathode region to the channel and did not include field fluctuations. Marini and Pakter \cite{marini2017single} subsequently examined the dynamics of single electrons in simplified static electromagnetic fields using a Hamiltonian formulation. They showed a transition from regular to chaotic trajectories while preserving the overall confinement bounds. Such chaotic motion enlarges the sampled phase space region and the effective ionization volume, suggesting that nonlinear single particle dynamics may contribute to ionization and anomalous cross field transport.

In the present work, we perform a test particle study of electron transport in a Hall thruster using field data extracted from the EDI resolving 3D PIC simulation described in Ref.~\cite{liu2026nearwallpathwaysanomalouselectron}. Four primary electric field configurations are considered. First, the full PIC electric field is used as the reference case, retaining the complete spatiotemporal structure of the EDI induced fluctuations. Second, the averaged field provides a static mean field case in which the EDI induced fluctuations are removed. Third, the POD field is used to identify the dominant EDI related field structures and evaluate their ability to reproduce anomalous electron transport. Fourth, a simplified analytical $E_y$ field is introduced to capture the azimuthal fluctuation structure observed in the full PIC field and to examine separately the effects of fluctuation amplitude and frequency on electron transport. In addition, a magnetic-field-only case and a denoised electric field case are included as controls.
For each case, transport is characterized using six diagnostics: the retained fraction, mean axial displacement $\langle\Delta z\rangle$, cumulative channel-entering fraction, residence time in the channel, final mean kinetic energy, and boundary resolved loss fraction. Together, these diagnostics clarify how EDI associated electric field fluctuations promote anomalous cross field electron transport in Hall thrusters.

The remainder of the paper is organized as follows. Sec.~\ref{sec:sim_model_setup} describes the test particle model and simulation setup. Sec.~\ref{sec:electric_field_models} introduces the electric field configurations considered in this study. Sec.~\ref{sec:results_and_analysis} analyzes the resulting electron transport, evaluates the reconstructed fields, and discusses the influence of azimuthal fluctuations.  Conclusions and discussion are presented in Sec.~\ref{sec:conclusions_and_discussion}.

\section{Test Particle Model and Simulation Setup}
\label{sec:sim_model_setup}

\subsection{Test Particle Simulation Framework}
\label{sec:framework}

\begin{figure}[htbp]
    \centering
    \includegraphics[width=0.9\linewidth]{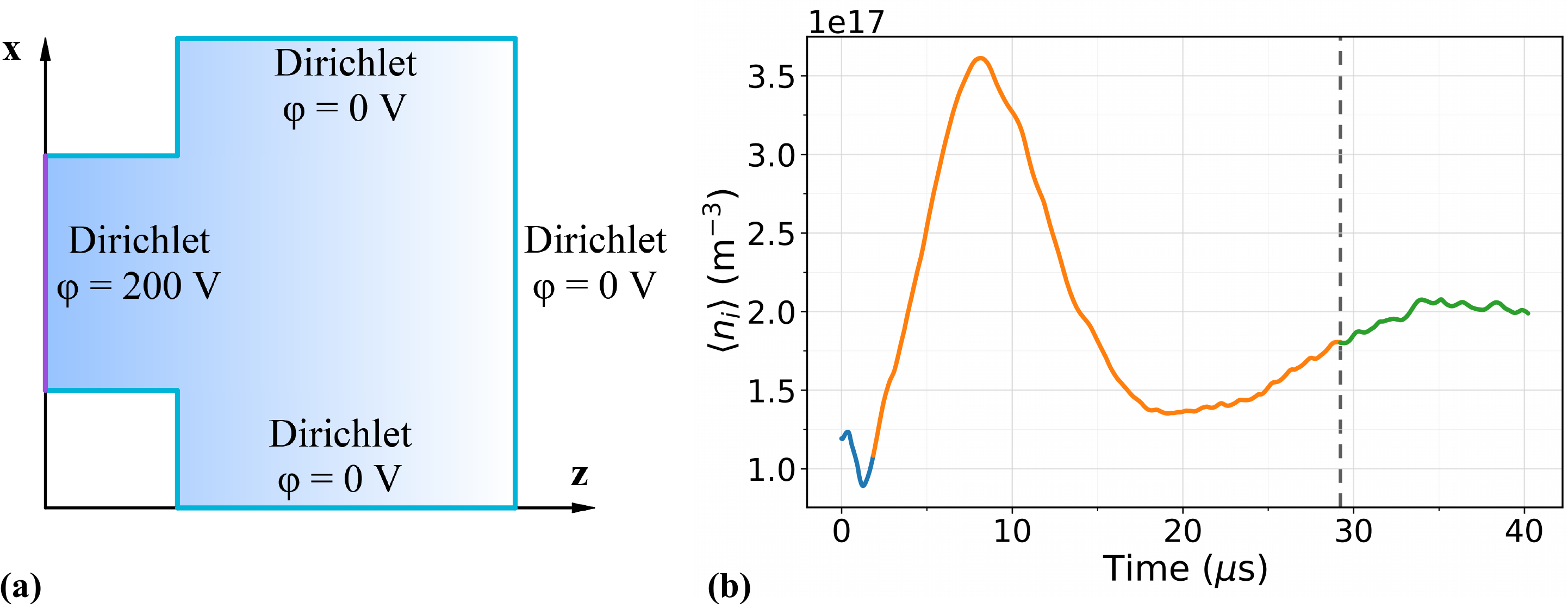}
    \caption{
    Potential boundary and ion density evolution for Case D \cite{liu2026nearwallpathwaysanomalouselectron}.
(a) Dirichlet potential boundary conditions.
(b) Time evolution of the spatially averaged ion number density, $\langle n_i\rangle$.
    }
\label{fig:caseD}
\end{figure}

As shown in Fig.~\ref{fig:caseD}, the test particle study is based on Case D of Ref.~\cite{liu2026nearwallpathwaysanomalouselectron}, in which Dirichlet electrostatic boundary conditions are applied on all non periodic boundaries: the anode face is held at $200~\mathrm{V}$, while all other outer boundaries are grounded. Fig.~\ref{fig:caseD}(b) shows the corresponding time evolution of the spatially averaged ion number density, $\langle n_i \rangle$. The green segment indicates the selected time interval from $t=29~\mu\mathrm{s}$, when $\langle n_i\rangle$ has approached a quasi-steady state, to $t=40.22~\mu\mathrm{s}$. 
Over this interval, the electric field data obtained from the 3D PIC simulation are then used as the primary dataset for the test particle study.

Electron test particles are advanced independently in the prescribed fields, without mutual interactions.
Ionization, neutral dynamics, and sheath interaction 
are not explicitly modeled for the test particles;
their collective effects are included only indirectly through the PIC derived electric fields. To isolate the role of the electric field, all numerical and geometrical
settings are fixed across cases and inherited from the 3D PIC setup
wherever applicable. The prescribed electric field is the only varied component;
its different configurations are introduced in
Sec.~\ref{sec:electric_field_models}.

The numerical simulations in the present work are performed using AlgoPlasma, an in-house plasma-simulation library formerly referred to as PMSL \cite{chen2025influence,ZHONG2026131809,zhao2025effect,chen2025coupling,Zhou_2025} and currently being prepared for a future open-source release. A central design goal of AlgoPlasma is to make the numerical algorithms used in plasma simulations explicit, modular, and testable. Specifically, key procedures such as Poisson solving, particle collisions, neutral dynamics, and particle pushing are implemented as separable components, allowing different simulation workflows to be assembled and executed as needed.

\subsection{Computational Domain and Magnetic Field}

\begin{figure*}[htbp]
    \centering
    \includegraphics[width=0.9\linewidth]{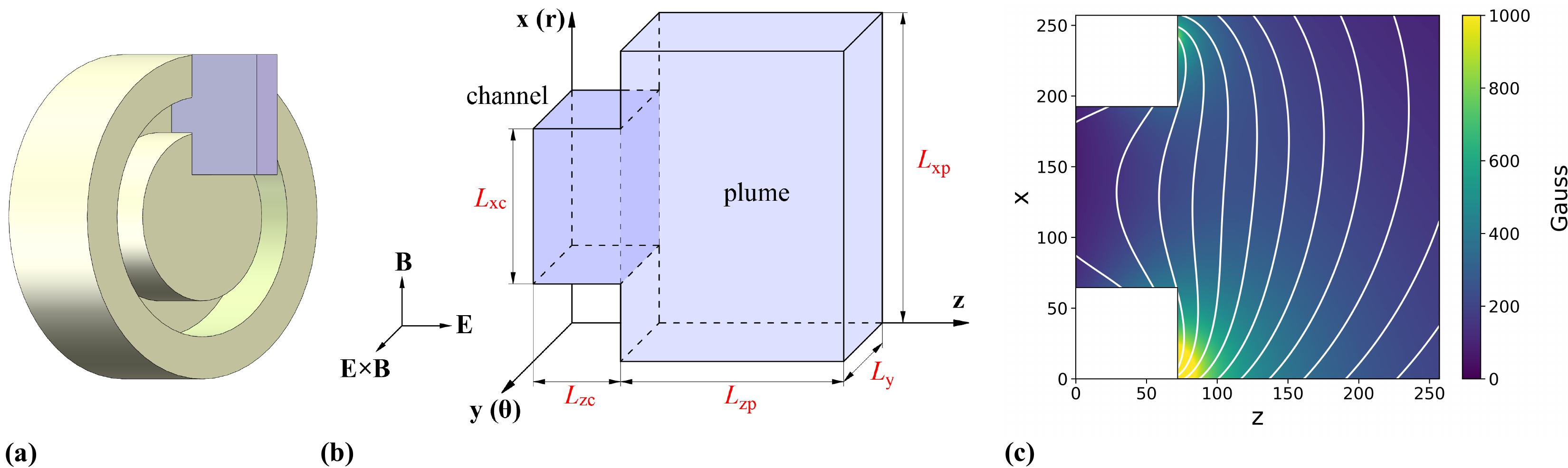}
    \caption{
Computational domain and magnetic field configuration.
(a) Schematic of the annular Hall thruster and the reduced azimuthal sector.
(b) Three dimensional domain in the local coordinate system $(x,y,z)\equiv(r,\theta,z)$, including the channel region (periwinkle) and plume region (lavender), with the nominal $\mathbf{E}$, $\mathbf{B}$, and $\mathbf{E}\!\times\!\mathbf{B}$ directions indicated.
(c) Magnetic field distribution in the $x$--$z$ plane, where colors denote $|\mathbf{B}|$ in Gauss and white contours denote isolines of the magnetic vector potential. The axis labels indicate grid indices rather than physical coordinates; in particular, $x=0$ does not correspond to $r=0$.
}
    \label{fig:simulation_domain}
\end{figure*}

The computational domain employed in this work is illustrated in Fig.~\ref{fig:simulation_domain}(a)(b).
An azimuthally reduced sector of an annular Hall thruster is modeled in a three-dimensional Cartesian coordinate system $(x,y,z)$,
which is locally aligned with the radial, azimuthal, and axial directions $(r,\theta,z)$, respectively.
Consequently, the azimuthal curvature of the full annulus is neglected,
as commonly done in previous PIC studies~\cite{chen2025influence,villafana20233d,Villafana_2021}.

The computational domain consists of two connected regions:
the discharge channel and the downstream plume region.
Detailed geometrical dimensions are summarized in Tab.~\ref{tab:numerical_para}. The entire domain is discretized using a uniform Cartesian mesh with
$\Delta x=\Delta y=\Delta z=0.1~\mathrm{mm}$.
The background Cartesian box contains
$N_x\times N_y\times N_z=256\times64\times256$ cells.
The actual domain is constructed by removing two rectangular cuboidal blocks of the upper left and lower left corners.

\begin{table}[htbp]
\centering
\caption{Details of the computational domain.}
\label{tab:numerical_para}
\small
\setlength{\tabcolsep}{6pt}
\begin{tabular*}{\linewidth}{@{\extracolsep{\fill}}l c c c@{}}
\toprule
\textbf{Dimension} & \textbf{Symbol} & \textbf{Value}~($\mathrm{mm}$) & \textbf{Cells}\\
\midrule
channel radial width   & $L_{xc}$ & 12.8  & 128\\
channel axial length   & $L_{zc}$ & 7.2   & 72\\
plume axial length     & $L_{zp}$ & 18.4  & 184\\
plume radial extent    & $L_{xp}$ & 25.6  & 256 \\
azimuthal width        & $L_y$    & 6.4   & 64\\
\bottomrule
\end{tabular*}
\end{table}

The imposed magnetic field is obtained from
the magnetic-circuit design of a Hall thruster prototype.
An axisymmetric magnetostatic model is constructed using
Finite Element Method Magnetics (FEMM).
Therefore, the magnetic field consists only of radial and axial components,
which correspond to the $x$ and $z$ directions in the local Cartesian
coordinate system. No azimuthal magnetic field component is included.
The magnetic field distribution exported from FEMM is mapped onto the
computational $x$--$z$ plane, as shown in Fig.~\ref{fig:simulation_domain}(c).
It is then extended uniformly along the azimuthal direction by assuming
${\partial \mathbf{B}} / {\partial y}=0$
yielding $\mathbf{B}(x,y,z)=[B_x(x,z),0,B_z(x,z)]$ in the three dimensional domain.

\subsection{Particle Advancement and Boundary Treatment}

The motion of electrons is advanced by solving the non-relativistic Lorentz force equation. 
For an electron with charge $q_e=-e$ and mass $m_e$, where $e$ is the elementary charge, the equations of motion are
\begin{equation}
    \frac{\mathrm{d} \mathbf{x}_p}{\mathrm{d}t} = \mathbf{v}_p ,
\end{equation}
and
\begin{equation}
    \frac{\mathrm{d}\mathbf{v}_p}{\mathrm{d}t}
    =
    -\frac{e}{m_e}
    \left[
        \mathbf{E}(\mathbf{x}_p,t)
        +
        \mathbf{v}_p \times \mathbf{B}(\mathbf{x}_p)
    \right],
\end{equation}
where $\mathbf{x}_p$ and $\mathbf{v}_p$ denote the position and velocity of the electron. 
The electromagnetic fields $\mathbf{E}(\mathbf{x}_p,t)$ and $\mathbf{B}(\mathbf{x}_p)$ are evaluated at the particle position. 

Since the electromagnetic fields are defined on the computational mesh, 
the electric field at the particle position $\mathbf{x}_p$ is obtained by 
trilinear interpolation over the eight neighboring cell centers:
\begin{equation}
    \mathbf{E}_p
    =
    \sum_{i=1}^8
    \mathbf{E}_{i}
    W(\mathbf{x}_p-\mathbf{x}_{i}),
\end{equation}
where $i$ labels the eight neighboring cell centers, $\mathbf{x}_{i}$ is the corresponding cell center position, and $W$ is the interpolation weight. 
In this work, the multidimensional weight is written as
\begin{equation}
    W(\mathbf{x}_p-\mathbf{x}_{i})
    =
    \prod_{\alpha=x,y,z}
    \left[
    1-
    \left|
        \frac{x_{p,\alpha}-x_{i,\alpha}}{\Delta x_\alpha}
    \right|
    \right] .
\end{equation}
The magnetic field $\mathbf{B}_p$ is obtained in the same manner.

Particle velocities are advanced using the Boris pusher, while particle positions are updated explicitly in the leapfrog scheme. 
The scheme is second-order accurate in time and preserves the phase space volume of charged particle motion. 
A fixed time step of $\Delta t = 5\times 10^{-12}\,\mathrm{s}$ is adopted, 
matching that used in the PIC simulation.

Periodic boundary conditions are imposed on the two $x$--$z$ boundary planes, which are normal to the $y$ direction. 
When a particle leaves the computational domain through one of these two boundaries, it is reintroduced from the opposite boundary while its velocity remains unchanged. 
All other boundaries are treated as absorbing boundaries. 
If a particle is detected outside the computational domain through any non-periodic boundary after the position update, it is immediately removed from the simulation. 

\subsection{Initial Particle Distribution and Injection Protocol}

\begin{figure}[htbp]
    \centering
    \includegraphics[width=0.9\linewidth]{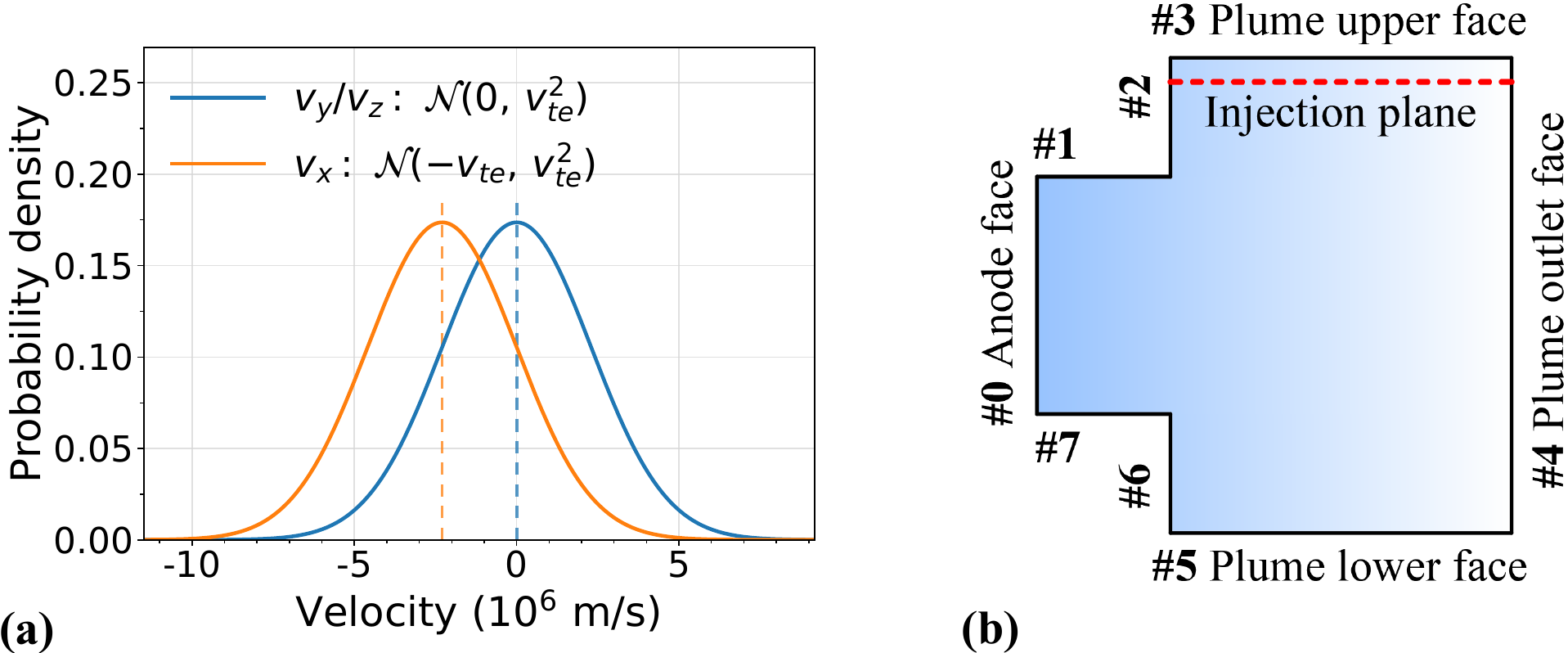}
    \caption{
    Initial velocity and spatial distribution of injected electrons.
    (a) Theoretical drifting Maxwellian velocity distribution for injected electrons. 
    (b) Particle injection plane and boundary nomenclature.
    }
\label{fig:init_velocity}
\end{figure}

Electrons are initialized through a cathode injection model. To maintain consistency with the PIC simulations, the electron temperature is set to $T_e=30~\mathrm{eV}$, which determines the one dimensional thermal velocity $v_{te}=\sqrt{k_B T_e/m_e}$.
Each injected electron is assigned a velocity sampled from a drifting Maxwellian distribution. 
The drift is imposed only on the $x$-velocity component, while the $y$- and $z$-velocity components are centered at zero. 
Thus, with $\mathbf{v}=(v_x,v_y,v_z)^\top$, the velocity distribution can be written as
\begin{equation}
    \mathbf{v} \sim \mathcal{N}\left(\mathbf{v}_d,\,v_{te}^{2}\mathbf{I}\right),
\end{equation}
where $\mathbf{v}_d=(-v_{te},0,0)^\top$.
The corresponding probability density functions are shown in Fig.~\ref{fig:init_velocity}(a).
The negative drift in the $x$ direction imposes a net electron flux from the cathode into the computational domain, while the isotropic thermal spread represents the finite electron temperature.

Spatially, the injection plane is defined at the radial grid index $x_{\mathrm{c}}=246$, as shown in Fig.~\ref{fig:init_velocity}(b). 
Injected electrons are sampled uniformly over the $y$--$z$ extent of this plane, while their initial $x$-position is assigned as
$x = x_{\mathrm{c}} + v_x \Delta t$,
corresponding to the displacement over one time step due to the injection velocity.
Particles outside the computational domain are rejected and resampled.

Regarding the timing and number of particle injection, two protocols are implemented in this work: single batch injection and population controlled injection.
In the single batch injection protocol, $N_0$ electrons are introduced only at the start of the simulation, with no further injection thereafter. In the population controlled injection protocol, electrons are injected only when the number of electrons in the computational domain, $N(t)$, falls below a prescribed target value, $N_{\mathrm{tar}}$, so as to maintain an approximately constant particle population. The main test particle parameters are summarized in Tab.~\ref{tab:test_particle_parameters}.

\begin{table}[htbp]
\centering
\caption{Main test particle simulation parameters.}
\label{tab:test_particle_parameters}
\small
\setlength{\tabcolsep}{6pt}
\begin{tabular*}{\linewidth}{@{\extracolsep{\fill}}l c c c@{}}
\toprule
\textbf{Quantity} & \textbf{Symbol} & \textbf{Value} \\
\midrule
time step & $\Delta t$ & $5\times10^{-12}~\mathrm{s}$\\
single batch injection number & $N_0$ & $10000$\\
target particle number & $N_{\mathrm{tar}}$ & $10000$\\
electron temperature    & $T_e$ & $30~\mathrm{eV}$\\
total integration steps & -- & $100000$\\
total duration & $t_{\mathrm{sim}}$ & $5\times10^{-7}~\mathrm{s}$\\
\bottomrule
\end{tabular*}
\end{table}

\subsection{Transport Diagnostics and Statistical Analysis}

To evaluate electron transport under different electric-field models, all test particle cases are initialized from the same electron distribution and analyzed using a common set of diagnostics. 
Let $N(t)$ be the number of active particles,
$N_{\rm inj}(t)$ the cumulative number of injected particles,
$\mathcal{A}(t)$ the active particle set, and
$\mathcal{I}(t)$ the injected particles set.
The survival fraction for single batch injection and the retained fraction
for population controlled injection are defined as
\begin{equation}
S(t)=\frac{N(t)}{N_0},
\qquad
R(t)=\frac{N(t)}{N_{\rm inj}(t)} .
\end{equation}
The mean axial displacement of active particles is
\begin{equation}
\langle\Delta z(t)\rangle
=
\frac{1}{N(t)}
\sum_{p\in\mathcal{A}(t)}
\left[z_p(t)-z_{p,0}\right],
\end{equation}
where $z_{p,0}$ is the injection position of particle $p$.
The cumulative channel-entering fraction is
\begin{equation}
F_{\rm ch}(t)=\frac{N_{\rm ch}(t)}{N_{\rm inj}(t)} ,
\end{equation}
where $N_{\rm ch}(t)$ is the number of injected particles that have entered
the channel at least once by time $t$. The channel is defined by
$z\le L_{zc}=7.2~\mathrm{mm}$, corresponding to grid index $72$.
The cumulative residence time of particle $p$ in the channel is
\begin{equation}
\tau_{{\rm ch},p}(t)=
\int_{t_{p,0}}^{\min(t,t_{p,\rm end})}
I_{\rm ch}\left[z_p(t)\right] \,\mathrm{d}t,
\end{equation}
where $t_{p,\rm end}$ is the loss time or the final simulation time, and
$I_{\rm ch}=1$ inside the channel and zero otherwise. The mean cumulative
residence time is
\begin{equation}
\langle\tau_{\rm ch}(t)\rangle
=
\frac{1}{N_{\rm inj}(t)}
\sum_{p\in\mathcal{I}(t)}
\tau_{{\rm ch},p}(t).
\end{equation}
Boundary losses are reported as
\begin{equation}
F_b(t)=\frac{N_b^{\rm loss}(t)}{N_{\rm inj}(t)},
\end{equation}
with $b$ denoting each absorbing boundary. Particle energization, representative trajectories, and force components are also examined.

These diagnostics quantify how EDI induced electric field fluctuations contribute to anomalous electron cross field transport, assess whether reduced field reconstructions reproduce the reference PIC field behavior, and examine the sensitivity of electron transport to prescribed fluctuation amplitude and frequency.

\section{Hierarchical Electric Field Modeling}
\label{sec:electric_field_models}

\begin{figure*}[htbp]
    \centering
    \includegraphics[width=0.95\linewidth]{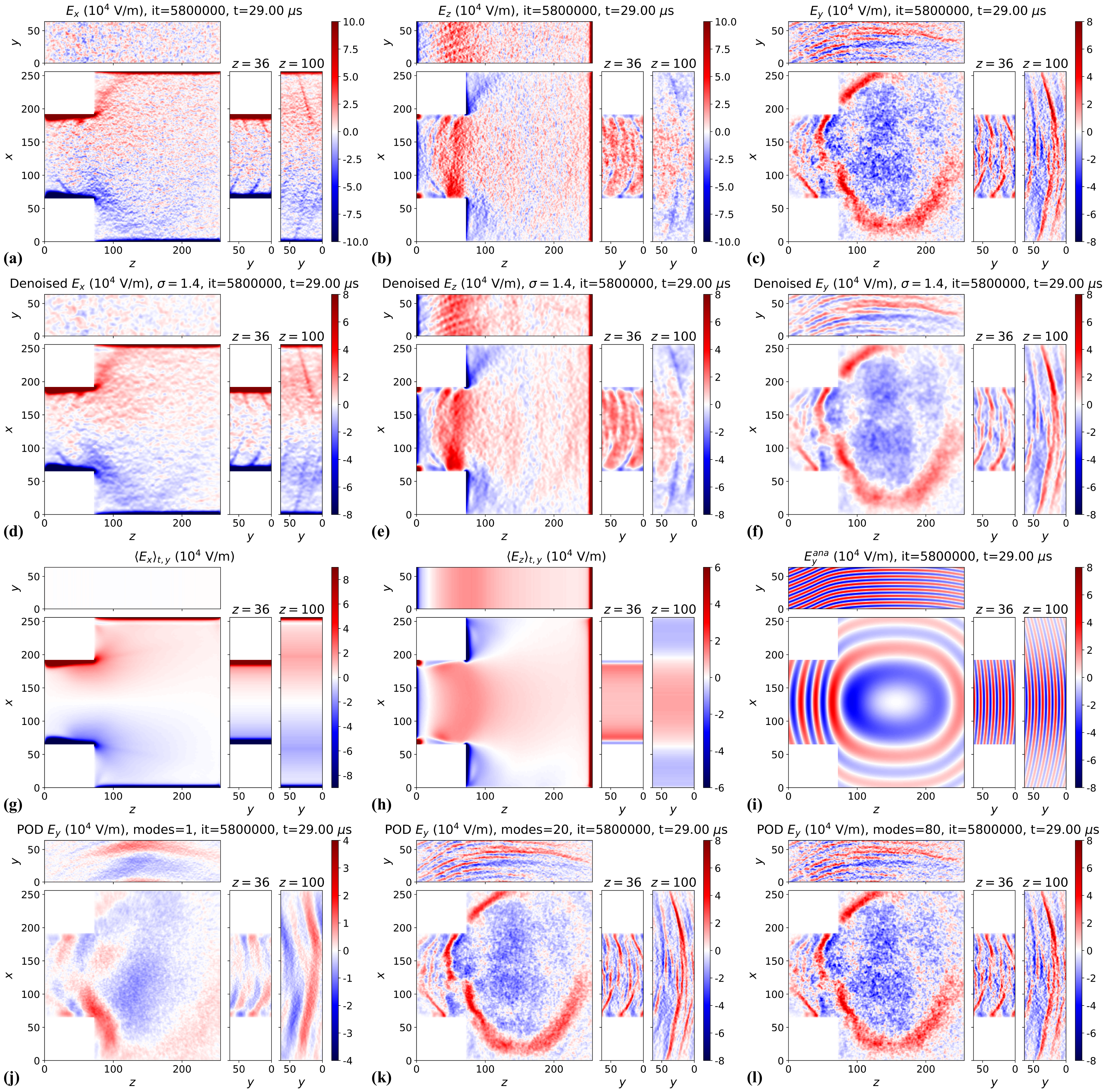}
    \caption{
    Spatial distributions of the electric field components at time step 
    $it=580000$ ($t=29~\mu\mathrm{s}$). 
    Each panel contains the mid-plane $z$--$y$ and $z$--$x$ cuts, together with 
    $y$--$x$ slices at grid indices $k_z=36$ inside the channel and $k_z=100$ in the near exit plume region. 
    Panels (a)--(c) show the full PIC fields ordered as $E_x$, $E_z$, and $E_y$. 
    Panels (d)--(f) show the corresponding denoised fields obtained by Gaussian smoothing with $\sigma=1.4$ grid cells. 
    Panels (g) and (h) show the averaged fields $\langle E_x\rangle_{t,y}$ and $\langle E_z\rangle_{t,y}$. 
    Panel (i) shows the analytical azimuthal field $E_y^{\mathrm{ana}}$. 
    Panels (j)--(l) show POD reconstructed $E_y$ fields with $1$, $20$, and $80$ retained modes. 
    }
    \label{fig:electric_field}
\end{figure*}

A magnetic-field-only case with $\mathbf{E}=0$ is used as the baseline, isolating confinement and loss caused solely by the imposed magnetic field and particle injection. The complete set of field configurations is summarized in Tab.~\ref{tab:field_cases} and described below.

\begin{table}[htbp]
\centering
\caption{Summary of field cases used in the test particle simulations.}
\label{tab:field_cases}
\small
\setlength{\tabcolsep}{6pt}
\begin{tabular*}{\linewidth}{@{\extracolsep{\fill}}l l l c@{}}
\toprule
\textbf{Case} & \textbf{$E_x$, $E_z$} & \textbf{$E_y$} & \textbf{$t$ dep.}\\
\midrule
$\mathbf{B}$ only
& 0
& 0
& no \\

Averaged $\mathbf{E}$
& $\langle E_x\rangle_{t,y}$, $\langle E_z\rangle_{t,y}$
& $\langle E_y\rangle_{t,y}\simeq 0$
& no \\

PIC $\mathbf{E}$
& full PIC
& full PIC
& yes \\

Denoised $\mathbf{E}$
& smoothed PIC
& smoothed PIC
& yes \\

POD $\mathbf{E}$
& POD-R
& POD-R
& yes \\

Analytical $E_y$
& $\langle E_x\rangle_{t,y}$, $\langle E_z\rangle_{t,y}$
& analytical $E_y$
& yes \\
\bottomrule
\end{tabular*}
\end{table}

\subsection{Time Dependent PIC Electric Field}

As described in Sec.~\ref{sec:framework}, the primary reference electric field data are extracted from the PIC simulation and stored every 1000 time steps ($0.005~\mu\mathrm{s}$) to reduce computational and storage costs. A piecewise-linear temporal reconstruction is therefore employed in the test particle calculation. Specifically, for an intermediate time $t \in [t_n, t_{n+1})$, the electric field is evaluated as
\begin{equation}
\mathbf{E}_{\mathrm{PIC}}(\mathbf{x},t)
=
\mathbf{E}_{\mathrm{PIC}}^{n}(\mathbf{x})
+
\alpha
\left[
\mathbf{E}_{\mathrm{PIC}}^{n+1}(\mathbf{x})
-
\mathbf{E}_{\mathrm{PIC}}^{n}(\mathbf{x})
\right],
\end{equation}
where $\alpha = {(t-t_n)}/{(t_{n+1}-t_n)}$.
This interpolation is applied to the full vector field, yielding a continuous-in-time, piecewise-linear electric field for advancing the test particles.
The initial time snapshots of the three
electric field components are shown in Fig.~\ref{fig:electric_field}(a)--(c).

To isolate the influence of full PIC field noise on the test particle
dynamics, we construct a denoised counterpart. In this control case, the temporally interpolated field is further smoothed in space using a 3D Gaussian
filter:
\begin{equation}
\widetilde{\mathbf{E}}_{\mathrm{PIC}}(\mathbf{x},t)
=
\left(G_{\sigma} * \mathbf{E}_{\mathrm{PIC}}(\cdot,t)\right)(\mathbf{x}),
\end{equation}
where $G_{\sigma}$ is a Gaussian kernel, $*$ denotes
grid based spatial convolution, and $\sigma=1.4$ is the smoothing width measured in grid cells. The filtered
electric field components at the initial time are shown in
Fig.~\ref{fig:electric_field}(d)--(f).

\subsection{Temporally and Azimuthally Averaged Electric Field}
\label{sec:static_electric_field}

A temporally and azimuthally averaged field is constructed by averaging the stored PIC electric field snapshots over the selected time interval and along the periodic $y$-direction. This procedure removes both time dependent fluctuations and azimuthal variations, leaving a static field that is uniform in the azimuthal direction. The averaged field is defined as
\begin{equation}
{\mathbf{E}}_{\mathrm{AVG}}(x_i,y_j,z_k)
=
\frac{1}{N_t N_y}
\sum_{n=1}^{N_t}
\sum_{j=1}^{N_y}
\mathbf{E}_{\mathrm{PIC}}(x_i,y_{j},z_k,t_n).
\end{equation}
The resulting $\langle E_x\rangle _ {t,y}$ and $\langle E_z \rangle _ {t,y}$ distributions are shown in Figs.~\ref{fig:electric_field}(g) and (h), whereas the averaged azimuthal component $\langle E_y\rangle _ {t,y}$ is negligibly small.

\subsection{POD Reconstructed Electric Field}

Proper orthogonal decomposition (POD) \cite{TURBULENCE_POD,Holmes_Lumley_Berkooz_1996} is applied to the PIC electric field snapshots over the selected interval to construct a reduced order representation. After removing the temporal mean, the first 200 POD modes are computed separately for each electric field component. For $E_\alpha$ with $\alpha=x,y,z$, the truncated reconstruction is written as
\begin{equation}
E_{\alpha,\mathrm{POD}}(\mathbf{x},t)
=
\overline{E}_{\alpha}(\mathbf{x})
+
\sum_{m=1}^{r}
a_m^{(\alpha)}(t)\psi_m^{(\alpha)}(\mathbf{x}),
\end{equation}
where $\overline{E}_{\alpha}$ is the temporal mean field, 
$\psi_m^{(\alpha)}$ and $a_m^{(\alpha)}$ are the $m$th POD mode and its temporal coefficient, 
and $r$ is the truncation rank, satisfying $r\le 200$.

The cumulative POD energy fractions are summarized in Tab.~\ref{tab:pod_energy}. 
The spectra indicate that the fluctuation energies associated with $E_x$ and $E_z$ 
are distributed over a broader range of modes, whereas $E_y$ exhibits stronger 
energy compaction in the low order modes. 
Fig.~\ref{fig:electric_field}(j)--(l) shows the POD reconstructed $E_y$ fields 
obtained by retaining $r=1$, $r=20$, and $r=80$ modes, respectively. 
The $r=1$ reconstruction captures the dominant large scale structure, while 
the higher rank reconstructions progressively recover the spatial modulation 
and fine scale fluctuations.

\begin{table}[htbp]
\centering
\caption{Cumulative energy fractions normalized by the total energy contained in the first 200 POD modes.}
\label{tab:pod_energy}
\small
\setlength{\tabcolsep}{6pt}
\begin{tabular*}{\linewidth}{@{\extracolsep{\fill}}l c c c@{}}
\toprule
\textbf{Modes} & $E_x (\%)$ & $E_y (\%)$ & $E_z (\%)$\\
\midrule
1   & 15.53  & 13.67  & 11.45  \\
2   & 25.30  & 26.37  & 18.55  \\
4   & 34.29  & 36.44  & 29.09  \\
10  & 43.59  & 59.89  & 43.20  \\
20  & 51.68  & 74.55  & 52.87  \\
40  & 60.48  & 79.94  & 62.24  \\
60  & 66.68  & 83.04  & 68.58  \\
80  & 71.96  & 85.70  & 73.70  \\
200 & 100    & 100    & 100    \\
\bottomrule
\end{tabular*}
\end{table}

\subsection{Simplified Analytical Azimuthal Electric Field}

A simplified analytical representation of the azimuthal electric field is constructed as an envelope modulated traveling wave model with curved phase fronts,
\begin{equation}
E_{y}^{\mathrm{ana}}(\mathbf{x},t)
=
\alpha_A A(x,z)
\cos\left[
k_y\left(y-\varphi(x,z)-v_{\phi} t\right)+\phi_0
\right],
\end{equation}
where $k_y=2\pi/\lambda_y$ is the azimuthal wavenumber,
$v_{\phi}$ is the azimuthal phase velocity, and $\phi_0$ is the initial phase. $\alpha_A$ is the amplitude factor, initially set to 1. 
The amplitude envelope is prescribed as a uniform background fluctuation with a localized Gaussian enhancement,
\begin{equation}
A(x,z)
=
A_b
+
A_p
\exp\left[
-\frac{1}{2}
\left(
\frac{(x-x_0)^2}{\sigma_x^2}
+
\frac{(z-z_0)^2}{\sigma_z^2}
\right)
\right].
\label{eq:Ey_ana_amp}
\end{equation}
Here $A_b$ is the background amplitude, while $A_p$ controls the localized enhancement centered at $(x_0,z_0)$.
The function $\varphi(x,z)$ prescribes the spatial displacement of the wave crest in the azimuthal direction. It is constructed by smoothly sweeping an axial baseline curve along a radial circular arc. This construction provides a compact representation of the layered and curved wavefront topology observed in the full PIC electric field snapshots.

In this model, the azimuthal wavelength is set to
$
\lambda_y=1.0~\mathrm{mm},
$
corresponding to 6.4 wavelengths across the periodic azimuthal domain. The oscillation frequency is set to
$
f=5.5~\mathrm{MHz},
$
which gives the azimuthal phase velocity
$v_{\phi}=f\lambda_y$.
A representative snapshot of the analytical field is shown in Fig.~\ref{fig:electric_field}(i). The remaining components, $E_x$ and $E_z$, are prescribed by the averaged field described in Sec.~\ref{sec:static_electric_field}.

\section{Results and Analysis}
\label{sec:results_and_analysis}

\subsection{Transport Comparison of Electric Field Models}

\begin{figure*}[htbp]
    \centering
    \includegraphics[width=0.95\linewidth]{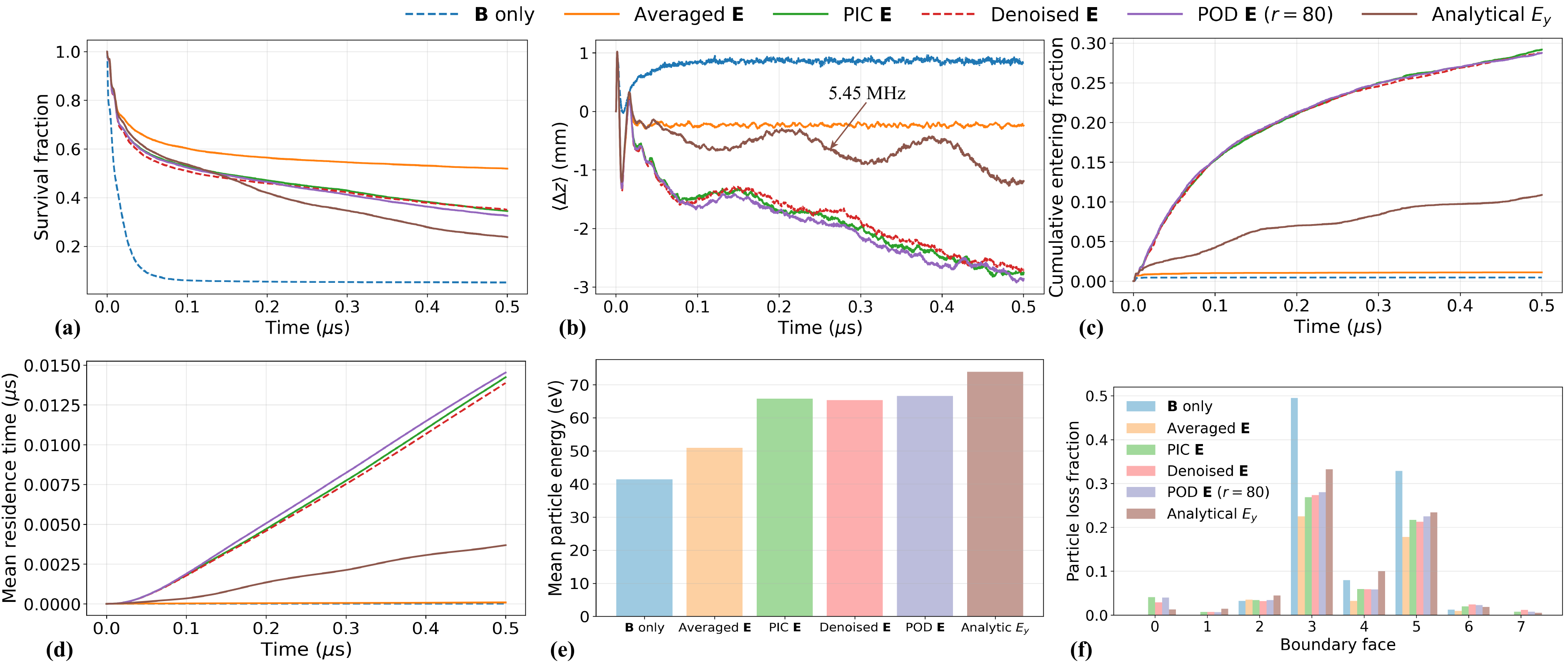}
\caption{
Single batch injection test particle statistics.
(a) Survival fraction $N(t)/N_0$.
(b) Mean axial displacement $\langle \Delta z\rangle$ of active particles.
(c) Cumulative channel-entering fraction.
(d) Mean cumulative residence time in the channel region.
(e) Mean kinetic energy of active particles at the final time.
(f) Boundary resolved loss fraction.
}
    \label{fig:single_injection}
\end{figure*}

\begin{figure*}[htbp]
    \centering
    \includegraphics[width=0.95\linewidth]{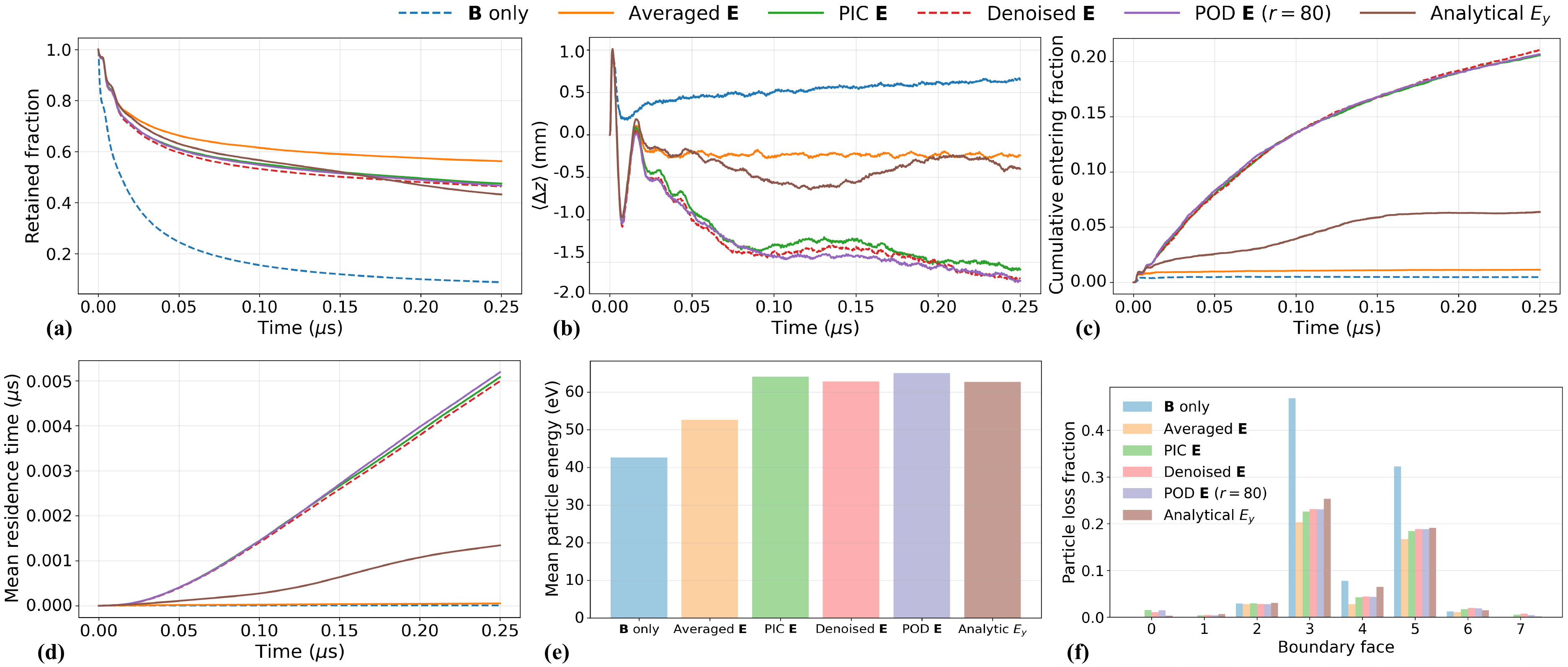}
\caption{
Population controlled injection test particle statistics.
(a) Retained fraction $N(t)/N_{\mathrm{inj}(t)}$.
(b) Mean axial displacement $\langle \Delta z\rangle$ of active particles.
(c) Cumulative channel-entering fraction.
(d) Mean cumulative residence time in the channel region.
(e) Mean kinetic energy of active particles at the final time.
(f) Boundary resolved loss fraction.
}
    \label{fig:maintain_injection}
\end{figure*}

Figs.~\ref{fig:single_injection} and \ref{fig:maintain_injection} compare the test particle statistics for single batch injection and population controlled injection, respectively. For each protocol, the six field cases summarized in Tab.~\ref{tab:field_cases} are evaluated using the same diagnostics: retained fraction, mean axial displacement $\langle\Delta z\rangle$, cumulative channel entering fraction, residence time in the channel region, mean kinetic energy of active particles at the final time, and boundary resolved loss fraction. Together, these quantities characterize particle confinement, axial transport, energization, and redistribution among loss channels.

Across all subpanels, the denoised field results closely match those obtained using the full PIC field, indicating that noise of the full PIC field has a negligible influence on the test particle analysis. Thus, the observed anomalous transport can be attributed to the intrinsic EDI electric field dynamics rather than to numerical noise.

In the single batch injection case, Fig.~\ref{fig:single_injection}(a) shows that particles are lost rapidly in the magnetic-field-only case, with only about $5.1\%$ of the initial particles remaining at the final time. The PIC and POD cases yield similar survival fractions, both around $35\%$, while the analytical $E_y$ cases are noticeably lower. The averaged field case gives the highest survival fraction, about $52\%$. However, this stronger retention does not imply stronger axial transport. The averaged field therefore improves confinement in the sense of particle retention, but it does not promote anode directed axial transport.

The distinction is more clearly shown in Fig.~\ref{fig:single_injection}(b). The magnetic-field-only and averaged field cases produce only weak axial displacement. Compared with the averaged field case, the analytical model, which introduces an azimuthal fluctuation, clearly enhances motion in the negative $z$ direction, yielding $\langle\Delta z\rangle\simeq -1.2~\mathrm{mm}$. This curve also exhibits a clear periodic signature, with a dominant frequency of about $5.45~\mathrm{MHz}$, in good agreement with the prescribed $5.5~\mathrm{MHz}$. Nevertheless, the axial transport produced by the analytical $E_y$ model remains only about half of that obtained with the PIC and POD fields, which reach $-3~\mathrm{mm}$.

A similar trend is reflected in the channel-entering and residence time statistics shown in Figs.~\ref{fig:single_injection}(c) and (d). The magnetic-field-only and averaged field cases form a weak transport group: their cumulative channel-entering fractions are only $0.45\%$ and $1.1\%$, and their mean residence times in the channel region are nearly zero. In contrast, the PIC and POD fields form a strong transport group. The POD reconstruction closely reproduces the PIC result, giving final channel-entering fractions of $28.5\%$, and mean residence times of $1.45\times10^{-2}~\mu\mathrm{s}$. The analytical $E_y$ case lies between these two groups, with a channel-entering fraction of about $10.9\%$ and a mean residence time of $3.7\times10^{-3}~\mu\mathrm{s}$.

Fig.~\ref{fig:single_injection}(e) shows that the magnetic-field-only and averaged field cases have lower final mean particle energies, about $41.4$ and $51.0~\mathrm{eV}$. By contrast, the PIC, POD, and analytical $E_y$ fields increase the mean energy to about $65.8$, $66.6$, and $73.9~\mathrm{eV}$, respectively.
Fig.~\ref{fig:single_injection}(f) shows that a large fraction of particle losses is concentrated on the plume upper and lower faces (3 and 5), indicating that magnetic field guided loss remains dominant. Most importantly, anode face loss occurs only in the PIC, POD, and analytical $E_y$ cases, with fractions of $4.1\%$, $4.0\%$, and $1.3\%$, whereas it is absent in the magnetic-field-only and averaged field cases. The averaged field case also has the smallest loss fractions on most faces, consistent with its highest survival fraction; this reduced boundary loss is related to the boundary sheath structure shown in Figs.~\ref{fig:electric_field}(g) and (h).

A consistent hierarchy is observed in the population controlled injection case, as shown in Fig.~\ref{fig:maintain_injection}. The population controlled injection case confirms that the hierarchy observed in the single batch injection case is not a transient consequence of the initially injected ensemble. Because particles are replenished after losses, the statistics emphasize sustained transport under each field model.

\subsection{Electron Trajectory Characteristics under the PIC Electric Field}

The statistical diagnostics demonstrate enhanced axial transport under the PIC field. To connect these ensemble level trends to microscopic dynamics, we next examine representative trajectories and force histories.

\begin{figure*}[htbp]
    \centering
    \includegraphics[width=0.9\linewidth]{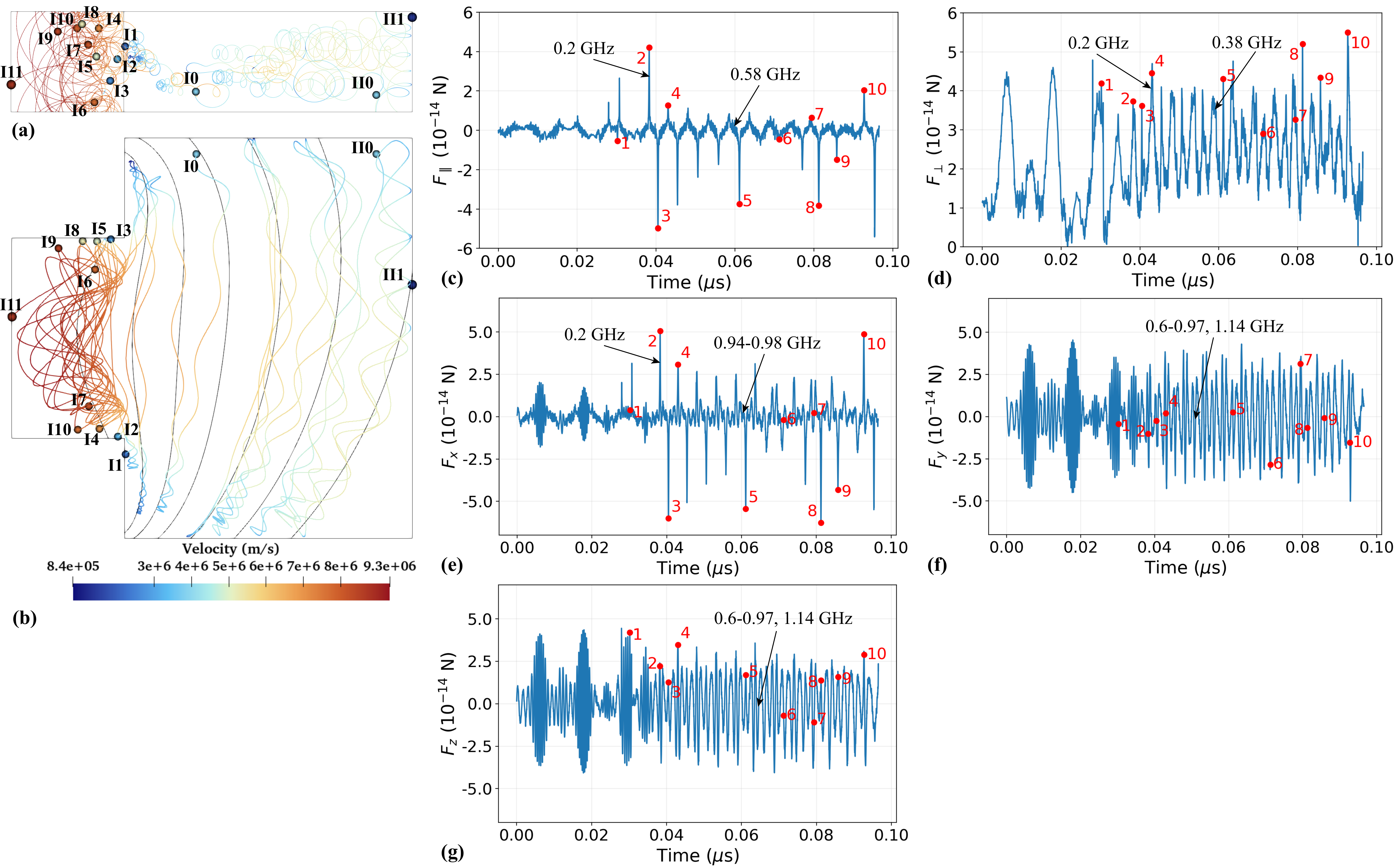}
    \caption{
    Electron trajectories and force histories under the full PIC electric field.
    (a) $z$--$y$ projection of particles I and II.
    (b) $z$--$x$ projection of particles I and II, colored by velocity magnitude.
    (c) Force component on particle I parallel to $\mathbf{B}$.
    (d) Magnitude of the force component perpendicular to $\mathbf{B}$.
    (e)--(g) Cartesian force components $F_x$, $F_y$, and $F_z$.
    The red markers 1--10 in (c)--(g) correspond to the same time instants as the numbered positions I1--I10 of particle I in (a) and (b).
    }
    \label{fig:traj_force}
\end{figure*}

Figs.~\ref{fig:traj_force}(a) and (b) compare two representative electron trajectories. Particle I is eventually transported to the anode face, whereas particle II leaves the computational domain through the plume outlet face and enters the plume neutralization region. The trajectories are characterized by radial magnetic-mirror motion and azimuthal Larmor rotation. The electron velocity reaches a maximum of approximately $9.3\times10^{6}~\mathrm{m/s}$ and is generally higher inside the discharge channel than in the plume. For particle II, the trajectory first shifts in the negative axial direction, then reverses before exiting the domain.

Figs.~\ref{fig:traj_force}(c)--(g) show the force history of particle I. The Lorentz force is decomposed into the component parallel to the magnetic field, $F_{\parallel}$, the perpendicular magnitude, $F_{\perp}$, and the Cartesian components $F_x$, $F_y$, and $F_z$. Pronounced peaks in $F_{\parallel}$ appear when the electron approaches the inner channel wall, where the near wall sheath electric field reflects the particle.

\begin{table}[htbp]
\centering
\caption{
Force statistics for the representative electron trajectory I in the PIC electric field case.
}
\label{tab:force_statistics}
\small
\setlength{\tabcolsep}{6pt}
\begin{tabular*}{\linewidth}{@{\extracolsep{\fill}}l c c c@{}}
\toprule
\textbf{Force} & \textbf{RMS}~($\mathrm{N}$) & \textbf{f}~($\mathrm{GHz}$) & \textbf{Range}~($\mathrm{N}$)\\
\midrule
$F_{\parallel}$ & $5.12\times10^{-15}$ & $0.2,\ 0.58$ & $-5.41\thicksim 4.21\times10^{-14}$ \\
$F_{\perp}$ & $2.45\times10^{-14}$ & $0.2,\ 0.38$ & $0.2\thicksim 5.5\times10^{-14}$ \\
$F_x$ & $9.0\times10^{-15}$ & $0.2,\ 0.94$-$0.98$ & $-6.27\thicksim 5.07\times10^{-14}$ \\
$F_y$ & $1.71\times10^{-14}$ & $0.6$-$0.97, \ 1.14$ & $-5.02\thicksim 4.54\times10^{-14}$ \\
$F_z$ & $1.59\times10^{-14}$ & $0.6$-$0.97,\ 1.14$ & $-4.08\thicksim 4.44\times10^{-14}$ \\
\bottomrule
\end{tabular*}
\end{table}

The corresponding force statistics are summarized in Tab.~\ref{tab:force_statistics}. The RMS value of the total Lorentz force is approximately $2.50\times10^{-14}~\mathrm{N}$. The RMS value of $F_{\perp}$ is $2.45\times10^{-14}~\mathrm{N}$, about $97.9\%$ of the total force RMS, whereas $F_{\parallel}$ is only about $20.4\%$. Thus, in the RMS sense, the force acting on particle I is dominated by the component perpendicular to $\mathbf{B}$.

The Lorentz force is further separated into the electric term $q\mathbf{E}$ and the magnetic term $q\mathbf{v}\times\mathbf{B}$. Their RMS magnitudes are $7.05\times10^{-15}~\mathrm{N}$ and $2.42\times10^{-14}~\mathrm{N}$, respectively. The magnetic term therefore provides the main contribution to the force amplitude, especially in the $y$ and $z$ directions. Although smaller in magnitude, the electric term remains important because it performs work on the electron and contributes to the particle energy variation and axial displacement.

The spectral results further distinguish the low frequency bounce motion from the high frequency gyromotion. The dominant peaks of $F_y$ and $F_z$ lie mainly in the range $0.6$--$1.1~\mathrm{GHz}$. In contrast, $F_x$ contains both a low frequency component near $0.2~\mathrm{GHz}$ and a higher frequency component near $0.95~\mathrm{GHz}$. The low frequency component is consistent with the dominant frequency of $F_{\parallel}$ and is associated with magnetic-mirror bounce motion in the discharge channel, whereas the peak near $0.95~\mathrm{GHz}$ corresponds to electron gyromotion.

\subsection{Effect of POD Truncation Order on Electron Transport}

\begin{figure*}[htbp]
    \centering
    \includegraphics[width=0.95\linewidth]{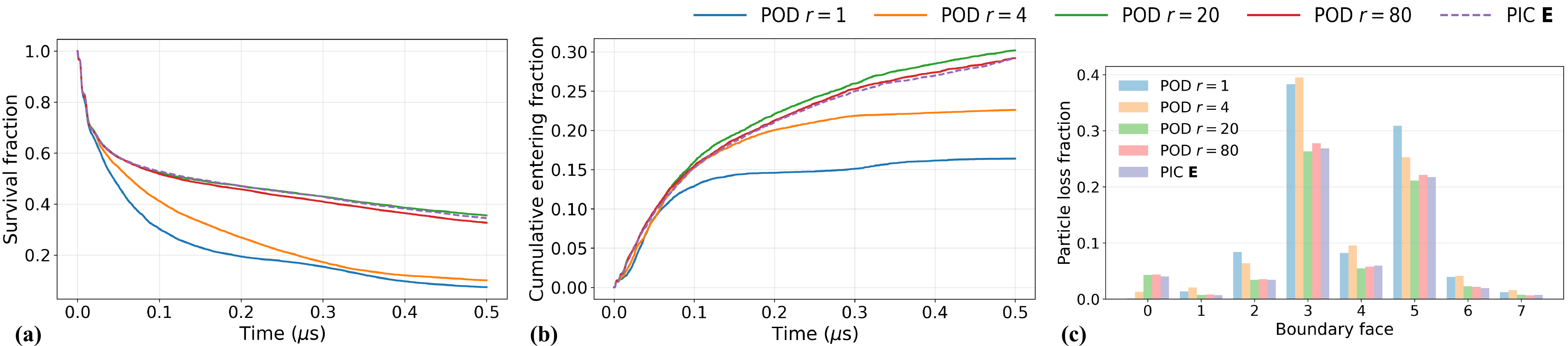}
\caption{
Single batch injection test particle statistics for POD reconstructions with different truncation orders. 
The PIC $\mathbf{E}$ result is included as a reference.
(a) Survival fraction $N(t)/N_0$.
(b) Cumulative channel-entering fraction.
(c) Boundary resolved loss fraction.
}
    \label{fig:pod_truncation_transport}
\end{figure*}

Although the POD reconstruction with $r=80$ gives the closest agreement with the full PIC field reference, POD is used here not simply to maximize field reconstruction accuracy, but to identify the dominant EDI modes and test whether a reduced set of modes can reproduce the associated anomalous electron transport. To this end, Fig.~\ref{fig:pod_truncation_transport} compares test particle statistics from POD fields with $r=1$, $4$, $20$, and $80$ against those from the full PIC field reference.

Fig.~\ref{fig:pod_truncation_transport}(a) shows that the low order reconstructions with $r=1$ and $r=4$ fail to reproduce the survival behavior of the reference case. Increasing the truncation order to $r=20$ substantially improves the agreement, giving a final survival fraction of approximately 0.35, comparable to those obtained with $r=80$ and the PIC field. A similar trend is observed for the cumulative channel-entering fraction in Fig.~\ref{fig:pod_truncation_transport}(b). The $r=1$ reconstruction significantly underestimates the entering fraction, whereas the $r=4$ case improves the early time behavior but still deviates at later times. In contrast, the $r=20$ reconstruction nearly reproduces the main temporal evolution of the entering process, with only moderate differences from the $r=80$ and PIC field results.

The boundary loss fraction in Fig.~\ref{fig:pod_truncation_transport}(c) further shows that the low order reconstructions substantially underestimate electron transport toward the anode face. Nevertheless, the anode face loss contribution is not completely absent: the loss fraction increases from only about $0.1\%$ for $r=1$ to about $1.3\%$ for $r=4$. By comparison, the $r=20$ case provides a much improved prediction. Its anode face loss fraction reaches about $4.3\%$, which is comparable to that of the $r=80$ case and is even slightly higher than the PIC field result. Moreover, the $r=20$ reconstruction yields losses at the other boundary faces that are comparable to those obtained from the $r=80$ reconstruction and the PIC field cases.

These comparisons demonstrate that, for the diagnostics considered here, $r=20$ is sufficient to recover the main PIC field transport signatures. This truncation order preserves the transport relevant field content while avoiding the additional modal complexity of higher order reconstructions.

\subsection{Amplitude and Frequency Effects in the Analytical $E_y$ Model}

\begin{figure*}[htbp]
    \centering
    \includegraphics[width=0.95\linewidth]{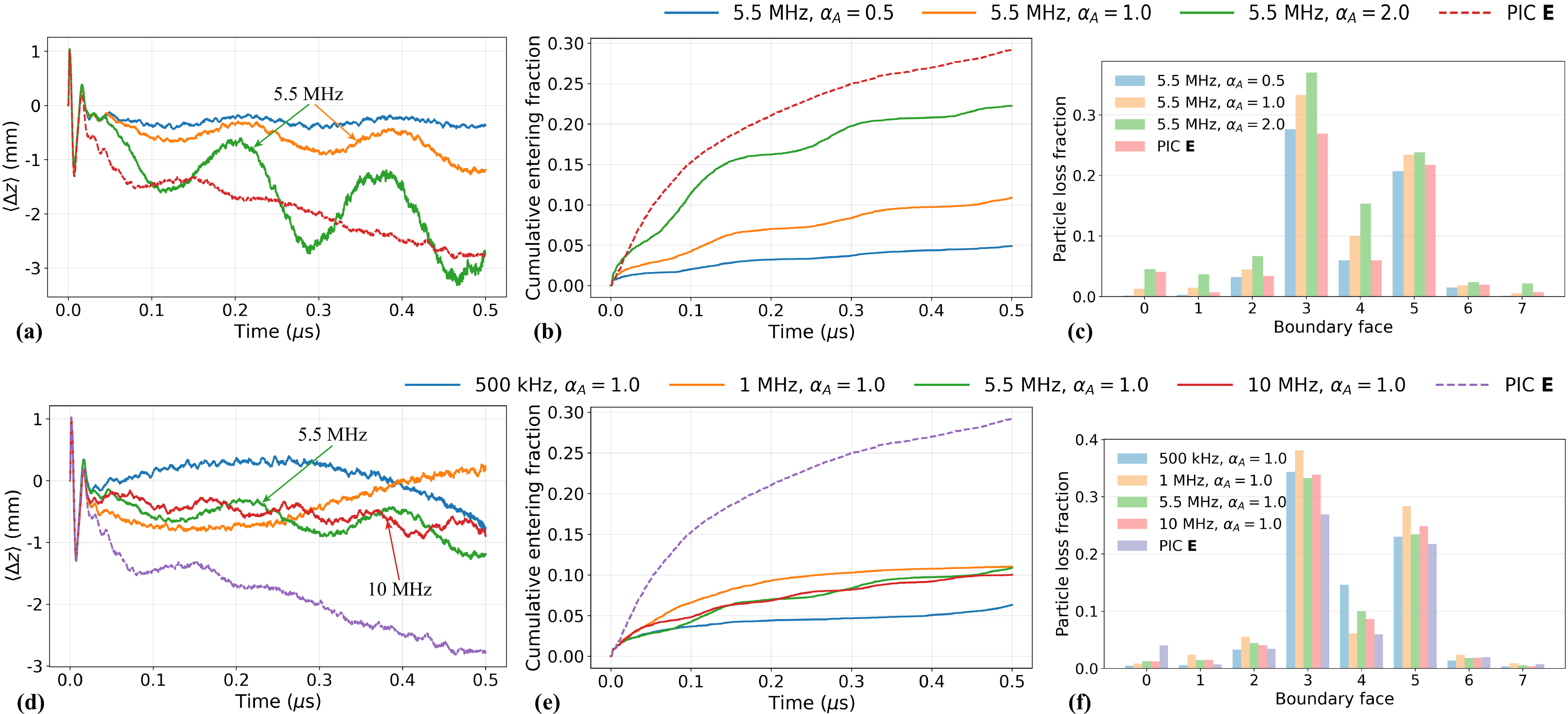}
\caption{
Effect of the prescribed analytical azimuthal electric field parameters on single batch injection electron transport.
The PIC $\mathbf{E}$ result is included as a reference.
(a)--(c) Amplitude scan at fixed frequency $f=5.5~\mathrm{MHz}$ with amplitude factor $\alpha_A=0.5$, $1$, and $2$.
(d)--(f) Frequency scan at fixed amplitude factor $\alpha_A=1$ with $f=500~\mathrm{kHz}$, $1~\mathrm{MHz}$, $5.5~\mathrm{MHz}$, and $10~\mathrm{MHz}$.
(a,d) Mean axial displacement $\langle \Delta z\rangle$.
(b,e) Cumulative channel-entering fraction.
(c,f) Boundary resolved loss fraction.
}
    \label{fig:scan_Ey}
\end{figure*}

Fig.~\ref{fig:scan_Ey} shows how the amplitude and frequency of the prescribed analytical $E_y$ field fluctuation affect electron transport under single batch injection, with the full PIC field case included as a reference.

As shown in Fig.~\ref{fig:scan_Ey}(a)--(c), at fixed frequency $f=5.5~\mathrm{MHz}$, increasing the amplitude factor from $\alpha_A=0.5$ to $2$ systematically enhances the axial response, although the overall transport level still does not fully match that of the PIC field case. The final mean axial displacement becomes increasingly negative, changing from approximately $-0.4~\mathrm{mm}$ for $\alpha_A=0.5$ to nearly $-3~\mathrm{mm}$ for $\alpha_A=2$. Meanwhile, the oscillatory signature in the $\langle\Delta z\rangle$ curves at the prescribed frequency becomes more pronounced as the fluctuation amplitude increases. At the same time, the cumulative channel-entering fraction increases from about $0.05$ to above $0.2$. The boundary resolved loss statistics show a consistent amplitude dependence: larger analytical field amplitudes generally lead to stronger boundary losses across the domain. Even in the weakest case, $\alpha_A=0.5$, approximately $0.1\%$ of the initially injected electrons still reach the anode face. These results indicate that, in the analytical model, the fluctuation amplitude directly controls the strength of the induced cross field transport.

As shown in Figs.~\ref{fig:scan_Ey}(d)--(f), the frequency scan at fixed amplitude factor $\alpha_A=1$ shows a weaker and less monotonic dependence than the amplitude scan. For the cumulative channel-entering fraction, the $500~\mathrm{kHz}$ case gives the weakest response, with a final value of approximately $0.065$, whereas the MHz-level cases yield comparable but distinguishable final fractions clustered around $0.10$. Nevertheless, all analytical field cases remain far below the response obtained with the PIC field.

The corresponding boundary resolved loss statistics are broadly consistent with this trend. The anode face loss fraction in the $500~\mathrm{kHz}$ case is significantly lower than those in the MHz-level cases, while all of them are still much smaller than that in the PIC case. For boundaries other than the plume outlet face, the $1~\mathrm{MHz}$ case generally gives larger loss fractions than the other analytical frequency cases, indicating that the frequency dependence is not strictly monotonic across different escape channels.

Thus, within the tested range, raising the frequency from $500~\mathrm{kHz}$ to the MHz range increases the response, but within the MHz range the dependence is weak and nonmonotonic. Overall, the analytical oscillatory $E_y$ component produces finite axial electron transport whose magnitude depends primarily on the fluctuation amplitude and more weakly on the fluctuation frequency.

\section{Conclusions and Discussion}
\label{sec:conclusions_and_discussion}

In the present work, electron transport in a Hall thruster was investigated using a
test particle approach driven by electric fields extracted from an EDI resolving
three dimensional PIC simulation. By comparing a hierarchy of field models, we
isolated the role of EDI induced electric field fluctuations in cross field
electron transport.
The denoised field case closely reproduces the full PIC field result for the
main transport diagnostics, indicating that numerical noise makes a negligible
contribution to the observed transport enhancement.

Comparisons among different field prescriptions show that the averaged field produces only weak axial transport, negligible channel residence time, and no anode face loss, despite retaining a finite axial electric field. In contrast, the full PIC field produces clear signatures of fluctuation driven transport, including pronounced negative axial displacement, substantial channel entry, finite channel residence, enhanced mean energy, and appreciable anode face loss. The EDI resolving field therefore does more than reduce confinement: it redirects a subset of magnetized electrons from plume dominated trajectories toward the channel and anode.

At the test particle level, trajectory resolved diagnostics provide a microscopic picture of this process. Representative trajectories show radial mirror-like motion and azimuthal gyromotion, together with intermittent axial displacements. The Lorentz force analysis shows that the instantaneous force is dominated by the component perpendicular to the magnetic field and by the magnetic part of the Lorentz force. Nevertheless, the electric field term remains essential because it performs work on the electrons and modifies their energy and axial motion. Near wall electric fields also produce pronounced parallel force peaks as particles approach the channel walls. These features connect the macroscopic transport statistics to gyromotion, mirror motion, wall interaction, and fluctuation driven energization.

The POD analysis shows that the transport relevant electric field content can be represented by a reduced set of modes. Although the POD modal energy is distributed over multiple modes, especially for $E_x$ and $E_z$, low truncation orders such as $r=1$ and $r=4$ do not fully reproduce the transport response. Increasing the truncation order to $r=20$ yields a marked improvement and recovers the main PIC field transport signatures, including anode directed losses comparable to those from the higher order reconstruction and the full PIC field. These results suggest that anomalous transport is largely governed by a finite set of modes whose combined spatiotemporal structure is required for transport. POD therefore provides both a reduced field representation and a diagnostic for identifying transport relevant fluctuations.

The analytical $E_y$ model further supports the role of azimuthal electric field fluctuations. Adding a prescribed oscillatory $E_y$ component to the averaged $E_x$ and $E_z$ fields produces finite axial displacement, channel entry, energization, and anode loss, which are nearly absent in the averaged field case. The amplitude scan shows that transport increases strongly with fluctuation amplitude, whereas the frequency scan shows a weaker and nonmonotonic dependence once the imposed frequency reaches the MHz range. However, the analytical field still does not fully reproduce the transport obtained with the full PIC fields. This discrepancy indicates that a single envelope modulated traveling wave captures only part of the mechanism; the full response also depends on multimode structure, three dimensional variation, and phase relations among electric field components.

Several limitations should be noted. Specifically, the present test particle model does not explicitly account for ionization, neutral dynamics, collisions, or detailed sheath interactions.
The analytical model also imposes a single coherent azimuthal wave and therefore cannot represent the broadband, multimode, and spatially nonuniform phase relations present in the PIC field. Future work should extend the framework to larger domains that better preserve magnetic field line connectivity and account for the curved channel geometry. Additional collisional and wall interaction processes should also be included, and the resulting effective electron transport should be compared with full PIC diagnostics and experimental measurements.

In summary, this test particle study demonstrates that EDI resolving three dimensional electric field fluctuations can strongly enhance electron transport in a Hall thruster. The averaged electric field alone cannot account for this transport, whereas POD fields with a moderate number of modes reproduce the main PIC field transport signatures. A simplified analytical azimuthal field captures part of the mechanism and shows that fluctuation amplitude is a primary determinant, but it remains insufficient to represent the full multimode EDI dynamics. These findings highlight the value of combining 3D PIC fields, reduced order reconstruction, and trajectory resolved test particle diagnostics for understanding anomalous electron transport in Hall thruster plasmas.

\section*{Acknowledgment}

The authors acknowledge the support from National Natural Science Foundation of China (Grant No. 52472403). This work was also partially supported by Harbin Institute of Technology Kunpeng \& Ascend Center of Cultivation.

\section*{Conflict of interest}
The authors have no conflicts to disclose.

\section*{Data Availability}

The data that support the findings of this study are available from
the corresponding author upon reasonable request.

%
%
%
%


\nocite{*}
\bibliography{reference}

\end{document}